\documentclass[12pt, onecolumn]{IEEEtran}
\usepackage{amsmath}
\usepackage{amssymb}
\usepackage{latexsym}
\newcommand{\EQ}{\begin{equation}}
\newcommand{\EN}{\end{equation}}
\newcommand{\w}{\mbox{wt}}

\newtheorem{theo}{Theorem}
\newtheorem{definition}{Definition}

\newtheorem{prop}{Proposition}
\newtheorem{lem}{Lemma}
\newtheorem{rem}{Remark}

\newcommand{\ba}{{\bf a}}
\newcommand{\bb}{{\bf b}}
\newcommand{\bg}{{\bf g}}

\newcommand{\bc}{{\bf c}}

\newcommand{\be}{{\bf e}}

\newcommand{\br}{{\bf r}}

\newcommand{\by}{{\bf y}}
\newcommand{\bx}{{\bf x}}

\newcommand{\bv}{{\bf v}}
\newcommand{\bu}{{\bf u}}

\newcommand{\bo}{{\bf 0}}

\newcommand{\Z}{\mathbb{Z}}
\newcommand{\F}{\mathbb{F}}

\newcommand{\al}{\alpha}

\newcommand{\dd}{\displaystyle}

\title{New completely regular
$q$-ary codes based on Kronecker products\thanks{This work has been partially supported by the Spanish MEC and the European FEDER Grants MTM2006-03250 and TSI2006-14005-C02-01 and also by the Russian fund of fundamental researches (the number of project 06 - 01 - 00226).
 Part of the material in Section~\ref{kro} was presented at the 2nd International Castle Meeting
on Coding Theory and Applications (2ICMCTA), Medina del Campo, Spain, September 2008.}}

\author{J. Rif\`{a}\thanks{Dept. of Information and Communications Engineering,
Universitat Aut\`{o}noma de Barcelona, 08193-Bellaterra, Spain}, V.A. Zinoviev
\thanks{Institute for Problems of Information
Transmission of the Russian Academy of Sciences, Bol'shoi Karetnyi
per. 19, GSP-4, Moscow, 101447, Russia}}
\date{v.8}

\begin{document}

\maketitle

\begin{abstract}

For any integer $\rho \geq 1$ and for any prime power $q$, the
explicit construction of a infinite family of completely regular
(and completely transitive)
$q$-ary codes with $d=3$ and with covering radius $\rho$ is given. The intersection array is also computed.
Under the same conditions, the explicit construction of an
infinite family of $q$-ary uniformly packed codes (in the wide
sense) with covering radius $\rho$, which are not completely
regular, is also given. In both constructions the Kronecker product
is the basic tool that has been used.
\end{abstract}

\section{Introduction}

Let $\F_q$ be a finite field of the order $q$.
Let $\w(\bv)$ denote the {\em Hamming weight} of a vector
$\bv \in \F_q^n$ and let $d(\bv, \bu)=\w(\bv-\bu)$ denote the
{\em Hamming distance} between two vectors $\bv,\bu \in \F_q^n$. We say that two vectors $\bv$ and $\bu$ are {\em neighbors} if $d(\bv, \bu)=1$.
A
$q$-ary linear $[n,k,d]_q$-code $C$ is a $k$-dimensional subspace of
$\F_q^n$, where $n$ is the {\em length}, $N=q^k$ is the {\em
cardinality} of $C$ and $d$ is the {\em minimum
distance},
\[
d~=~\min\{d(\bv, \bu):~\bv, \bu \in C,~\bv \neq \bu\}.
\]  The error correcting capability of a code
$C$ with minimum distance $d$ is given by $e=\lfloor (d-1)/2\rfloor$.

Given any vector $\bv \in \F_q^n$, its {\em distance to the code
$C$} is $d(\bv,C)=\min_{\bx \in C}\{ d(\bv, \bx)\}$
and the {\em covering radius} of the code $C$ is
$$
\rho=\max_{\bv \in \F_q^n} \{d(\bv, C)\}.
$$

Let $~D=C+\bx~$ be a {\em coset} of  $C$, where $+$ means the
component-wise addition in $\F_q$. The {\em weight} $\w(D)$ of $D$
is the minimum weight of the codewords of $D$. For an arbitrary
coset $D$ of $C$ of weight $s=\w(D)$ denote by $\mu(D) =
(\mu_0(D), \mu_1(D), ... , \mu_n(D))$ its weight distribution,
where $\mu_j(D)$, ~$j=0, \ldots, n$ denotes the number of words of
$D$ of weight $j$. Notice that $\mu_j(D)=0$ for all $j < s$.

\begin{definition}\label{de:1.1}
A $q$-ary linear code $C$ with covering radius $\rho$ is called
{\em completely regular} if the weight distribution of any coset $D$
of $C$ of weight $i$, $i=0,1,...,\rho$ is uniquely defined by the
minimum weight of $D$, i.e. by the number $i=\w(D)$.
\end{definition}

\begin{definition}\label{de:1.2}
Let $C$ be a $q$-ary code of length $n$ and let $\rho$ be its
covering radius. We say that $C$ is {\em uniformly packed}
in the wide sense, i.e. in the sense of \cite{Bas1}, if there exist
rational numbers
$~\al_0,~\ldots,~\al_{\rho}~$ such that for any $\bv~\in~\F_q^n~$
\EQ\label{eq:2.1}
\sum_{k=0}^{\rho}\al_k\,f_k(\bv)~=~1~,
\EN
where $f_k(\bv)$ is the number of codewords at distance $k$ from
$\bv$.
\end{definition}

The case $\rho=e+1$ corresponds to {\em uniformly packed codes},
suggested in \cite{Goe2}, and the case $\rho=e+1$ and
$\al_{\rho-1} =\al_{\rho}$ corresponds to {\em uniformly packed
codes in the narrow sense} or sometimes called {\em strongly
uniformly packed codes}, suggested in \cite{Sem2}; see more special
cases of such codes in \cite{Cohe, Goe1, Goe2, Sem2}.
It is well known (see, for example, \cite{Bro1}) that any
completely regular code is uniformly packed in the wide sense.
In turn, uniformly packed codes with $\rho=e+1$ are completely
regular \cite{Goe2, Sem2}, including some extended such codes
\cite{Bas1, Bas2, Sem2}.
But till now, the only known examples of uniformly packed codes,
which are not completely regular, were the known binary (primitive
in narrow sense) BCH codes of length $n=2^m-1$~ ($m$ odd) with
minimum distance $d=7$ \cite{Char} and the $\Z_4$-linear
Goethals-like codes of length $n=2^m-1$~ ($m$ even) with
minimum distance $d=7$ \cite{Hell} (including extended codes
for both families of codes). In both cases the codes have
covering radius $\rho = e+2 = 5$, and $\rho = e+3 = 6$ for
extended codes.

It has been
conjectured for a long time that if $C$ is a completely regular code
and $|C|>2$, then $e \leq 3$. For the
special case of  linear completely
transitive codes \cite{Sole}, the analogous conjecture was solved in~\cite{Bor1} and~\cite{Bor2} proving that for
$e \geq 4$ such nontrivial codes do not exist. Hence, the existing completely regular codes and completely transitive codes have an small error correcting capability. In respect of the covering radius,
Sol\'{e} in \cite{Sole} uses the direct sum of $\ell$ copies
of fixed perfect binary $1$-code of length $n$ to construct infinite families of binary completely regular codes of length
$n{\cdot} \ell$ with covering radius $\rho =\ell$. Thus, using~\cite{Sole}, the covering radius of the resulting code is growing
to infinity with the length of the code.

One of the main purpose in the current paper is to describe a method
of constructing linear completely regular and completely transitive
codes with arbitrary covering radius, which is constant when the length
of the resulting code is growing to infinity. More exactly, for
any prime power $q$ and for any natural number $\ell$ we give, in Theorem~\ref{theo:1},  an
explicit  construction of an infinite family of linear $q$-ary
completely regular and completely transitive codes with lengths
$n =(q^m-1)(q^{\ell}-1)/(q-1)^2$ and with fixed covering radius
$\rho = \ell$, where $m \geq \ell$ is any integer (a previous approach in this direction can be found in~\cite{RiZi}).
The intersection array for these completely regular codes is computed in Theorem~\ref{theo:4.1}.

Under the same conditions (i.e. for any prime power $q$ and for any
natural number $\ell$) we give the explicit construction of an
infinite family of $q$-ary linear uniformly packed codes (in the wide
sense) with lengths $n = (\ell+1)\,(q^m-1)/(q-1)$ and with
covering radius $\rho = \ell$, where $m,\ell \geq 2$ are any
integers. All these codes (with the exception $q=\ell=2$) are not
completely regular.

\section{Preliminary results}\label{dos}

For a given $q$-ary code $C$ with covering radius $\rho=\rho(C)$ define
\[
C(i)~=~\{\bx \in
\F_q^n:\;d(\bx,C)=i\},\;\;i=0,1,\ldots,\rho.
\]

We also use the following alternative standard definition of completely
regularity~\cite{Neum}.

\begin{definition}\label{de:2.2} A code $C$ is completely regular, if
for all $l\geq 0$ every vector $\bx \in C(l)$ has the same number
$c_l$ of neighbors in $C(l-1)$ and the same number $b_l$ of
neighbors in $C(l+1)$. Also, define $a_l = (q-1){\cdot}n-b_l-c_l$
and note that $c_0=b_\rho=0$. Refer to
$(b_0, \ldots, b_{\rho-1}; c_1,\ldots, c_{\rho})$ as the
intersection array of $C$.
\end{definition}


\medskip

For a $q$-ary $[n,k,d]_q$-code $C$ with weight distribution
$\mu(C) = (\mu_0, \ldots , \mu_n)$ define the {\em outer distance}
$s = s(C)$ as the number of nonzero coordinates $\mu^{\perp}_i$,~
$i=1, \ldots, n$ of the vector $(\mu^{\perp}_0, \ldots , \mu^{\perp}_n)$
obtained by the MacWilliams transform of $\mu(C)$ \cite{Dels}. Hence,
since $C$ is a linear code, $s(C)$ is the number of different nonzero
weights of codewords in the dual code $C^{\perp}$.

\begin{lem}[\cite{Dels}]\label{lem:2.2}
For any code $C$ with covering radius $\rho(C)$ and with outer distance
$s(C)$ we have $\rho(C) \leq s(C)$.
\end{lem}

\begin{lem}\label{lem:2.3}
Let $C$ be a code with minimum distance $d=2e+1$, covering radius $\rho$
and outer distance $s$. Then:
\begin{enumerate}
\item Code $C$ is uniformly packed in the wide sense
if and only if $\rho=s$ ({\cite{Bas2}}).
\item If $C$ is completely regular then it is uniformly
packed in the wide sense ({\cite{Bro1}}).
\item If $C$ is uniformly packed in the wide sense and
$\rho=e+1$, then it is completely regular ({\cite{Sem2,Goe2}}).
\end{enumerate}
\end{lem}

Let $C$ be a linear code of length $n$ over $\F_q$, a finite field of
size a prime power $q$. Following~\cite{Macw}, if $q=2$, the automorphism
group $Aut(C)$ of $C$ is a subgroup of the symmetric group $S_n$ consisting
of all $n!$ permutations of the $n$ coordinate positions which send $C$ into itself.

Let $M$ be a monomial matrix, i.e. a matrix with exactly one nonzero entry
in each row and column. If $q$ is prime, then $Aut(C)$ consists of all
$n\times n$ monomial matrices $M$ over $\F_q$ such that $\bc M \in C$
for all $\bc \in C$. If $q$ is a power of a prime number, then $Aut(C)$ also
contains all the field automorphisms of $\F_q$ which preserve $C$.

The group $Aut(C)$ induces an action on the set of cosets of $C$ in the
following way: for all $\phi\in Aut(C)$ and for every vector
$\bv \in \F_q^n$ we have $\phi(\bv + C) = \phi(\bv) + C$.

In~\cite{Sole} it was introduced the concept of completely transitive
binary linear code and it can be generalized to the following definition,
which also corresponds to the definition of coset-completely transitive code
in~\cite{giudici}.

\begin{definition}\label{ctc}
Let $C$ be a linear code over $\F_q$ with covering radius $\rho$. Then $C$ is completely
transitive if $Aut(C)$ has $\rho +1$ orbits when acts on the cosets of $C$.
\end{definition}

Since two cosets in the same orbit should have the same weight distribution, it is clear that
any completely transitive code is completely regular.

\section{Kronecker product construction}\label{kro}

In this section we describe a new construction which provides for
any natural number $\rho$ and for any prime power $q$ an infinite
family of $q$-ary linear completely regular codes with covering
radius $\rho$.

\begin{definition}
For two matrices $A=[a_{r,s}]$ and $B =[b_{i,j}]$ over
$\F_q$ define a new matrix $H$ which is the Kronecker product
$H = A \otimes B$, where $H$ is obtained by changing any element
$a_{r,s}$ in $A$ by the matrix $a_{r,s} B$.
\end{definition}

Consider the matrix $H = A \otimes B$ and let $C$, $C_A$ and $C_B$ be the codes over
$\F_q$ which have, respectively, $H$, $A$ and $B$ as a parity check matrices. Assume that $A$ and
$B$ have size $m_a \times n_a$ and $m_b \times n_b$, respectively. For
$r\in\{1,\cdots, m_a\}$ and $s\in \{1,\cdots, m_b\}$ the rows in $H$
look as
$$(a_{r,1}b_{s,1}, \cdots, a_{r,1}b_{s,n_b}, a_{r,2}b_{s,1}, \cdots,
a_{r,2}b_{s,n_b}, \cdots, a_{r,n_a}b_{s,1}, \cdots, a_{r,n_a}b_{s,n_b}).$$
Arrange these rows taking blocks of $n_b$ coordinates as columns
such that the vectors $\bc$ in code $C$ are presented as
matrices of size $n_b \times n_a$:
\EQ\label{equation1}\bc = \left [
\begin{array}{ccc}
c_{1,1} & \ldots & c_{1,n_a}\\
c_{2,1} & \ldots & c_{2,n_a}\\
\vdots &\vdots &\vdots \\
c_{n_b,1} & \ldots & c_{n_b,n_a}
\end{array}
\right ]
=
\left [
\begin{array}{ccc}
&\bc_1 & \\
&\bc_2 & \\
&\vdots &\\
&\bc_{n_b} &
\end{array}
\right ],
\EN
where $c_{i,j} = a_{r,j} b_{s,i}$ and $\bc_r$ denotes the $r$-th row
vector of this matrix.

We will call matrix representation the above way to present the vectors $\bc\in C$.

Let us go to a further view on the codewords of $C$, the code over $\F_q$
which has $H=A\otimes B$ as a parity check matrix.
Consider vector $\bc\in C$
and use the representation in~(\ref{equation1}), hence $\bc=(\bc_1, \bc_2,\cdots, \bc_{n_b})^t$, where $(\cdot)^t$ means the transpose vector. Now compute the syndrome vector which leads us to a $(m_b\times m_a)$ matrix that we will equal to zero. We have
\begin{equation}\label{syn}
B\big( A\bc_1^t,A\bc_2^t,\ldots,A\bc_{n_b}^t\big)^t =0\,\, \mbox{ and so, }
B\big(A\bc^t\big)^t=B{\cdot}\bc{\cdot}A^t=0.
\end{equation}

With this last property it is easy to note that any ($n_b\times n_a$) matrix with
codewords of $C_A$ as rows belong to the code $C$ and also any
($n_b\times n_a$) matrix with codewords of $C_B$ as columns
belongs to the code $C$. Vice versa, all the codewords in $C$ can
always be seen as linear combinations of matrices of both types above.

Moreover, it is straightforward to state the following well known fact.

\begin{lem}\label{lem:4.0}
Codes defined by the parity check matrices $A\otimes B$ and $B\otimes A$
are permutation equivalent.
\end{lem}

\medskip

From now on, we  assume that  matrix $A$ (respectively, $B$) is a
parity check matrix of a Hamming code with parameters $[n_a,k_a,3]_q$
(respectively, $[n_b,k_b,3]_q$), where $n_a = (q^{m_a}-1)/(q-1) \geq 3$
(respectively, $n_b=(q^{m_b}-1)/(q-1) \geq 3$) and
$k_a=n_a-m_a$ (respectively, $k_b=n_b-m_b$).

Denote by $H_m$ the parity check matrix of a perfect Hamming $[n,k,3]_q$-code
$C$ over $\F_q$, where $n=(q^m-1)/(q-1)$.
Let $\xi_0=0,\xi_1=1,\ldots,\xi_{q-1}$ denote the elements of
$\F_q$. Then the matrix $H_m$  can be expressed, up to equivalence,
through the matrix $H_{m-1}$ as follows \cite{Sem1}:
\[
H_m = \left[
\begin{array}{cc|c|c|c|c}
&0\cdots 0\,&\,1\cdots 1\,&\,\cdots\,&\,\xi_{q-1}\cdots
\xi_{q-1}\,&\,1\,\\\hline
&H_{m-1}\,&\,H_{m-1}\,&\,\cdots \,&\,H_{m-1}\,&\,{\bf 0}\,\\
\end{array}\right],
\]
where ${\bf 0}$ is the zero column and where $H_1 = [1]$.
 Note that, under such construction, the following lemmas are straightforward (see, for example, \cite{Sem1}).

 \begin{lem}
\label{i} Matrix $H_m$ contains as columns, among other, all the
  $m$ possible binary vectors of length $m$ and of weight $1$.
 \end{lem}

\begin{lem}\label{lem:4.1a} For $i=1,\ldots,m$, let $\br_i$ denote
the $i$-th row of $H_m$. Let $\bg = \sum_{i=1}^m \xi_i \br_i$, with
$\xi_i \in \F_q$, be any linear combination of the rows of $H_m$.
If $\w(\bg) \neq 0$, then $\w(\bg) = q^{m-1}$.
\end{lem}

Throughout this work we will consider the columns in $A$ and $B$ ordered in such a way that the one-weighted vectors will be placed in the first $m_a$ (respectively, $m_b$) positions.

Any codeword $\bc \in C$, which has nonzero elements only in one row
(or only in one column) will be called a {\it line}. Since  $A$
and $B$ are parity check matrices of Hamming codes (i.e. they have minimum
distances $3$), there are lines of weight $3$. For example, a row line
$L_r=(\alpha_1,\alpha_2,\alpha_3)_{(s_1,s_2,s_3)}$ (respectively, a column line
$L_s=(\alpha_1,\alpha_2,\alpha_3)_{(r_1, r_2, r_3)}$) means that the codeword
$\bc$ of weight $3$, whose nonzero $r$th row (respectively, nonzero $s$th column)
has nonzero elements $\alpha_1$, $\alpha_2, \alpha_3$ in columns $s_1$th, $s_2$th,
$s_3th$ (respectively, in rows $r_1$th,  $r_2$th, $r_3$th). Recall that this means the
following equality for the corresponding
columns $\ba_{s_1}$, $\ba_{s_2}$, and $\ba_{s_3}$ of matrix $A$ (respectively, for
the columns $\bb_{r_1}$, $\bb_{r_2}$, and $\bb_{r_3}$ of matrix $B$):
\EQ\label{eq:5}
\sum_{i=1}^3 \mu_i\,\ba_{s_i} ~=~\bo~~~(\mbox{respectively},
\sum_{j=1}^3 \lambda_j\,\bb_{r_j}~=~\bo).
\EN

Define the set of row indices as $R=\{1, \ldots, n_b\}$ (respectively, of
column indices as $S=\{1, \ldots, n_a\}$) and assume that the first $m_b$ indices (respectively, the first $m_a$) corresponds to the column vectors in $A$ (respectively, in $B$) of weight one.  By definition of perfect codes, for a
fixed row index $r\in R$ (respectively, column index $s\in S$), for any two
nonzero elements $\alpha_1,\alpha_2 \in \F_q$ and for any two different
$s_1, s_2 \in S$ (respectively, $r_1, r_2 \in R$) there is a unique row
line $L_r=(\alpha_1,\alpha_2,\alpha_3)_{(s_1, s_2, s_3)}$ (respectively,
column line $L_s=(\alpha_1,\alpha_2,\alpha_3)_{(r_1, r_2, r_3)}$) for some nonzero element
$\alpha_3 \in \F_q$ and for some $s_3 \in S$ (respectively,  $r_3 \in R$).

It is well known that the linear span of the vectors of weight three in
a Hamming code gives all the code. Hence, the linear span
of the row lines of weight three and the column lines of weight three
gives all the codewords of $C$.

Given a vector $\bv \in \F_q^{n_b{\cdot}n_a}$ let $\bv = [v_{ij}]$ be its matrix
representation.  We will call main submatrix the $(m_b\times m_a)$ matrix containing the first $m_b$ rows and $m_a$ columns of the matrix representation. It is easy to see that, after simplifying (i.e. passing lines through the points placed out of the main submatrix), we can obtain a new vector $\bv'$ in the same coset $\bv+C$ such that its matrix
representation has zero elements everywhere except into the main submatrix $M_{\bv}$.

\begin{lem}
Let $\bv \in \F_q^{n_b{\cdot}n_a}$ be a vector and let $M_{\bv}$ be its main submatrix representation. Then:
\begin{enumerate}
\item Vector $\bv$ is in $C$ if and only if $M_{\bv}=0$.
\item For each $\bv$ the main submatrix representation $M_{\bv}$ is unique.
\end{enumerate}
\end{lem}
\begin{proof}
First of all, take a nonzero $(m_b \times m_a)$ matrix $M$. Each column (respectively, row) is not a line, indeed, we would have a line $L_r=(\alpha_1,\ldots,\alpha_{m_a})_{s_1,\ldots,s_{m_a}}$ involving only independent vectors of weight one, which is impossible. Hence, the conclusion is that it is impossible that such a nonzero main submatrix $M$ is a codeword. Vice versa, given a vector $\bv \in C$ and doing the simplification operations described above we will obtain a zero main submatrix representation.

The second point is a corollary of the first one.
\end{proof}

Given a vector $\bv \in \F_q^{n_b{\cdot}n_a}$ let $\bv = [v_{ij}]$ be its matrix
representation. Compute the syndrome $S_{\bv}$ like in~(\ref{syn}) which is a $(m_b\times m_a)$ matrix. Note that adding $(n_a-m_a)$ zero columns and $(n_b-m_b)$ zero rows to this syndrome matrix we obtain the above main submatrix representation $M_{\bv}$ for $\bv$.

Hence, in other words:

\begin{lem}\label{syn1}
Given a vector $\bv \in \F_q^{n_b{\cdot}n_a}$ let $\bv = [v_{ij}]$ be its matrix
representation. Then: $$(A\otimes B)(\bv) = S_{\bv} = B [v_{ij}] A^t = B M_{\bv} A^t$$
\end{lem}

Consider $m_b \geq m_a$ (in the contrary case we will do the same but reverting the role of matrices $A$ and $B$). Take a vector $\be \in \F_q^{n_b{\cdot}n_a}$ such that all the elements in the matrix representation are zeroes, except one. So, there are two specific values $1\leq \lambda \leq n_b$, $1 \leq \mu \leq n_a$ such that $\be=[e_{ij}]$; $e_{\lambda \mu}=e$ and $e_{ij}=0$ for all $i\not= \lambda$ and $j\not= \mu$.

Using~(\ref{eq:5}), we can pass a column line across the point $(\lambda,\mu)$ obtaining one or more aligned points in the first $m_b$ rows. Again, passing row lines across these last points we obtain the main submatrix representation which is as follows:

\begin{equation}\label{one}
M_{\mu \otimes \lambda} = e{\cdot}\left[ \begin{array}{cccc}
\mu_1\lambda_1 &   \mu_2\lambda_1 & \cdots &\mu_{m_a}\lambda_1 \\
\mu_1\lambda_2 & \mu_2\lambda_2 & \cdots & \mu_{m_a}\lambda_2 \\
\vdots &\vdots &\vdots &\vdots \\
\mu_1\lambda_{m_b} & \mu_2\lambda_{m_b} & \cdots &\mu_{m_a}\lambda_{m_b}
\end{array}\right],
\end{equation}
where $\ba_{\mu}= \sum_{i=1}^{m_a} \mu_i \ba_{s_i}$; $\bb_{\lambda}= \sum_{i=1}^{m_b} \lambda_i \bb_{r_i}$ and $\ba_{s_i}$, $\bb_{r_i}$ are the  one weighted vectors of length $m_a$ and $m_b$, respectively.

\begin{rem}\label{index}
Note that the first nonzero indexes in $\{\mu_1,\mu_2,\cdots, \mu_{m_a}\}$ and $\{\lambda_1,\lambda_2,\cdots,\lambda_{m_b} \}$ are $\mu_{f_\mu} =1$ and $\lambda_{f_\lambda}=1$, respectively.
\end{rem}

It is important to point out that given a $(m_b\times m_a)$ matrix $M$ the $rank(M_{\mu \otimes \lambda}+M)$ differs from $rank(M)$ in one unit, at the most.

\begin{prop}\label{prop:dist}
Let $\bv \in \F_q^{n_b{\cdot}n_a}$ be a vector and $M_{\bv}$ be its main submatrix representation.
Then the distance
of $\bv$ to code $C$ is $d(\bv,C)=rank(M_{\bv})$.
\end{prop}

\begin{proof}
Let $rank(M_{\bv})=s$. Doing
simplifications passing lines across the rows of $M_{\bv}$ we will obtain a representation vector with nonzero elements in, at maximum, $s$ columns. Again passing lines across these columns we obtain a representation matrix for the given vector $\bv$ with not more that $s$ nonzero coordinates. Hence,  $d(\bv,C) \leq s$.

Now, we are going to prove that $s\leq d(\bv,C)$. Consider the vector $\bc \in C$ with
the same coordinates as $\bv$ and, moreover the new $d(\bv,C)$ coordinates that we need to add to $\bv$ to obtain that vector $\bc$ in $C$.

For each one of the coordinates $v_{ij}$ in which   $\bv$ and $\bc$ differ we do the same consideration as in~(\ref{one}) and so, we see that the rank of the main submatrix representation of $\bv + v_{ij}$ differs from the previous in one unit, at the most. That is, after adding all the necessary coordinates to $\bv$ to obtain $\bc$, the rank of the main submatrix representation varied in, at the most, $d(\bv,C)$ units obtaining the final value of zero. Hence, the initial rank $s$ must be necessarily less or equal to $d(\bv,C)$.
\end{proof}

The following theorem shows that the code constructed by the Kronecker
product is a completely transitive code and, therefore, is a completely regular
code.

\begin{theo}\label{theo:1}
Let $C$ be the code over $\F_q$
which has $H=A\otimes B$ as a parity check matrix, where $A$ and $B$ are
parity check matrices of Hamming codes $[n_a,k_a,3]_q$ and  $[n_b,k_b,3]_q$,
respectively, where $n_a = (q^{m_a}-1)/(q-1) \geq 3$; $n_b=(q^{m_b}-1)/(q-1) \geq 3$;
$k_a=n_a-m_a$ and $k_b=n_b-m_b$. Then:
\begin{enumerate}
\item Code $C$ has length $n=n_a{\cdot}n_b$,
dimension $k= n - m_a{\cdot}m_b$ and minimum distance $d = 3$.
\item The covering radius of $C$ is $\rho=min\{m_a,m_b\}$.
\item Code $C$ is completely transitive and, therefore, a completely
regular code.
\end{enumerate}
\end{theo}

\begin{proof}  It is straightforward to check that the code $C$ has length
$n=n_a{\cdot}n_b$, dimension $k= n - m_a{\cdot}m_b$ and minimum distance $d = 3$.

In respect of the covering radius, take a vector
$\bv \in \F_q^{n_b{\cdot}n_a}$ and use Proposition~\ref{prop:dist}. Matrix $M_{\bv}$ is a $(m_b\times m_a)$ matrix, so this rank is an integer value from $0$ to $min(m_a,m_b)$.

To prove that $C$ is a completely transitive code it is enough to show that
starting from two vectors $\bx, \by \in C(\ell)$, there exists a monomial
matrix $\phi\in Aut(C)$ such that $\phi(\bx) \in \by +C$ or, in other words, $(A\otimes B)(\phi(\bx)) = (A\otimes B)(\by)$.

First of all, let
$\phi_1$ be any monomial $(n_a\times n_a$) matrix and $\phi_2$ be any monomial
($n_b\times n_b$) matrix. It is clear that $$(A\phi_1)\otimes (B\phi_2) = (A\otimes B)(\phi_1 \otimes \phi_2)$$
and $\phi_1\otimes \phi_2$ is a monomial ($n_an_b\times n_an_b$) matrix.
Moreover, we have that $Aut(A \otimes B) = Aut\big((A\otimes B)^\perp\big)= Aut(C)$ and so $Id\otimes \phi \in Aut(C)$.
Hence, if $\phi$ is an automorphism in $Aut(B)$ then $Id\otimes \phi \in Aut(C)$.

The two given vectors $\bx, \by$ belong to $C(\ell)$ and so, $rank(S_{\bx}) =rank(S_{\by})$, where $S_x$ and $S_y$ are the syndrome of $\bx$ and $\by$, respectively. To prove that $C$ is a completely transitive code we will show that there exists an automorphism $\phi \in Aut(B)$ such that $(A\otimes B)(\by) = (A\otimes B\phi)(\bx) = (A\otimes B)(\phi(\bx))$.

Assume $m_b\geq m_a$ (otherwise, we will do the same construction reverting $A$ and $B$). It is straightforward to
find an invertible $(m_b\times m_b)$ matrix $K$ over $\F_q$ such that $S_{\bx}^t\,K=S_{\by}^t$.
Since $B$ is the parity check matrix of a Hamming code, the matrix $K^tB$ is again
a parity check matrix for a Hamming code and $K^tB=B\phi$ for some monomial matrix
$\phi$. Moreover, if $G_B$ is the corresponding generator matrix for this Hamming
code, i.e. $BG_B^t=0$, then $(B\phi)G_B^t= (K^tB)G_B^t=0$ and so $\phi \in Aut(B)$.

Finally,
$$
(A\otimes B)(\by) = S_{\by} = K^tS_{\bx} = K^t (B\bx A^t)= B\phi \bx A^t= (A\otimes B\phi)(\bx) =(A\otimes B)(\phi(\bx)).$$
\end{proof}

The following goal is to compute the intersection
array for this completely regular code $C$.

\begin{theo}\label{theo:4.1}
Let $C_A$ and $C_B$ be two Hamming codes of parameters $[n_a,k_a,3]_q$
and $[n_b,k_b,3]_q$, respectively, where $n_a = (q^{m_a}-1)/(q-1) \geq 3$;
$n_b=(q^{m_b}-1)/(q-1) \geq 3$ with dimension
$k_a=n_a-m_a$ and $k_b=n_b-m_b$, respectively. Let $A$ (respectively, $B$) be a
parity check matrix for the code $C_A$ (respectively, $C_B$).
Then the matrix $H = A \otimes B$, the Kronecker product of $A$
and $B$, is a parity check matrix of a $q$-ary completely regular
$[n, k, d]_q$-code $C$ with covering radius $\rho$, where
\EQ\label{eq:4.1}
n = n_a{\cdot}n_b,~~k = n - m_a{\cdot}m_b,~~ d = 3,~~
\rho = \min\{m_a,\,m_b\},
\EN
and with intersection numbers for $\ell=0,1, \ldots, \rho$:

$$
b_{\ell}=(q-1) \left(n_a-\frac{q^{\ell}-1}{q-1}\right)\left(n_b-\frac{q^{\ell}-1}{q-1}\right),
$$
$$
c_{\ell}=\frac{q^{\ell}-1}{q-1}q^{\ell-1},
$$
$$
a_\ell = (q-1){\cdot}n_a{\cdot}n_b -c_\ell -b_\ell
$$
\end{theo}

\begin{proof}
Let $\bx \in C(\ell)$ and $\by = \bx + \be$, where $\be$ is a
$(n_b \times n_a)$ matrix which has one nonzero position, say $e_{\lambda,\mu}$,
where $\lambda \in R$, $\mu \in S$ and
$e_{\lambda,\mu} = e \in \F_q^*$.

As we said before, after doing simplifications we can always think that $\bx=[x_{i,j}]$ has $\ell$ non zero positions at the main diagonal of value $1$ and the corresponding $\ell$ column vectors $\bb_i$ (respectively $\ba_j$) are linear independent. Let $R_1$ be the set of these column vectors $\bb_i$ (respectively, let $S_1$ be the set of these column vectors $\ba_j$).

The case $\ell=0$ follows immediately (any location of $e_{\lambda,\mu}$
contributes clearly only to the number $b_0$):
\[
a_0=0,~b_0=(q-1){\cdot}n_a{\cdot}n_b.
\]

Now consider the general case: $1 \leq \ell \leq \rho$.

First of all, assume that $\be=[e_{\lambda,\mu}]$ and that the vector $\bb_\lambda$ is linearly independent from the vectors in $R_1$ (respectively, $\ba_\mu$ is linearly independent from the vectors in $S_1$). The only contribution is $b_{\ell}$ and so it is easy to find that:
$$
b_{\ell}=(q-1) \left(n_a-\frac{q^{\ell}-1}{q-1}\right)\left(n_b-\frac{q^{\ell}-1}{q-1}\right).
$$

Now, we are going to the case where $\ba_\mu$ linearly depends from the set $S_1$ and $\bb_\lambda$ from $R_1$. From~(\ref{one}) and Proposition~\ref{prop:dist} we can assume that $\ba_\mu$ is linearly dependent from the vectors in $S_1$ and also $\bb_\lambda$ from the vectors in $R_1$. So, $\ba_\mu= \sum_{i=1}^{m_a} \mu_i \ba_{\mu_i}$; $\bb_\lambda= \sum_{j=1}^{m_b} \lambda_j \bb_{\lambda_j}$ and $\ba_{\mu_i}$, $\bb_{\lambda_j}$ are the  one weighted vectors of length $m_a$ and $m_b$, respectively.

We want to count in how many ways the following matrix has rank $\ell-1$:
\begin{equation}
Id_{\ell} + M_{\mu \otimes \lambda} =\left[ \begin{array}{cccccc}
1+\mu_1\lambda_1e &   \mu_2\lambda_1e & \cdots &  \mu_{\ell}\lambda_1e & \cdots &\mu_{m_a}\lambda_1e \\
\mu_1\lambda_2e & 1+\mu_2\lambda_2e & \cdots &  \mu_{\ell}\lambda_2e & \cdots &\mu_{m_a}\lambda_2e \\
\vdots &\vdots &\vdots &\vdots &\vdots &\vdots \\
\mu_1\lambda_{\ell}e & \mu_2\lambda_{\ell}e & \cdots &  1+\mu_{\ell}\lambda_{\ell}e & \cdots & \mu_{m_a}\lambda_{\ell}e \\
\mu_1\lambda_{\ell + 1}e & \mu_2 \lambda_{\ell +1}e & \cdots & \mu_{\ell}\lambda_{\ell + 1}e & \cdots & \mu_{m_a}\lambda_{\ell + 1}e \\
\vdots &\vdots &\vdots &\vdots &\vdots &\vdots \\
\mu_1\lambda_{m_b}e & \mu_2\lambda_{m_b}e & \cdots &\mu_{\ell}\lambda_{m_b}e & \cdots &\mu_{m_a}\lambda_{m_b}e
\end{array}\right]
\end{equation}

If $\lambda_{f_\lambda}=\mu_{f_\mu}=1$ (see Remark~\ref{index}) are such that $f_\lambda > \ell$ or $f_\mu >\ell$ then the rank of the above matrix $Id_{\ell} + M_{\mu \otimes \lambda}$ would be greater than $\ell-1$. Hence, we can transform the above matrix in the following one, which has the same rank:
\begin{equation}
\left[ \begin{array}{cccccc}
1+\sum_{i=1}^\ell \mu_i\lambda_ie &   \mu_2\lambda_1e & \cdots &  \mu_{\ell}\lambda_{1}e & \cdots &\mu_{m_a}\lambda_1e \\
0 & 1 &\cdots & 0& \cdots &0 \\
\vdots &\vdots &\vdots &\vdots &\vdots &\vdots \\
0& 0 & \cdots &  1 & \cdots & 0 \\
-\lambda_{\ell +1}e/\lambda_1 & 0 & \cdots & 0 & \cdots & 0\\
\vdots &\vdots &\vdots &\vdots &\vdots &\vdots \\
-\lambda_{m_b}/\lambda_1 e& 0&\cdots &0 & \cdots &0
\end{array}\right]
\end{equation}

It is easy to see that the rank of the above matrix is $\ell-1$ plus the rank of $P$, where

\begin{equation}
P=\left[ \begin{array}{cccccc}
1+\sum_{i=1}^\ell \mu_i\lambda_i e&    \mu_{\ell + 1}\lambda_1e & \cdots &\mu_{m_a}\lambda_1e \\
-\lambda_{\ell +1}e/\lambda_1 &  0 & \cdots & 0\\
\vdots &\vdots  &\vdots &\vdots \\
-\lambda_{m_b}e/\lambda_1  &0 & \cdots &0
\end{array}\right]
\end{equation}

We are interested in to count in how many ways the rank of $P$ is zero. It happens when all the $\mu_i=0$ for $\ell+1 \leq i \leq m_a$; $\lambda_i=0$ for $\ell+1 \leq i \leq m_b$ and $1+\sum_{i=1}^\ell \mu_i\lambda_ie =0$.

Now, note that when we fix a specific values for $\lambda_1, \lambda_2, \cdots, \lambda_{\ell}$, with $\lambda_{f_\lambda} =1$ we want to count how many solutions $(\mu_1,\mu_2,\cdots,\mu_{m_a})$ has the equation $1+\sum_{i=1}^\ell \mu_i\lambda_ie =0$ with the restrictions $\lambda_{f_\lambda}=\mu_{f_\mu}=1$ (see Remark~\ref{index}).

We know that $e$ is any value in $\F_q^{*}$ and, on the other side, $\mu_{f_\mu}=1$. Hence, each solution $(\mu_1,\mu_2,\cdots,\mu_{m_a})$ to the equation $1+\sum_{i=1}^\ell \mu_i\lambda_ie =0$ with the quoted restrictions could be transformed in to solution of $1+\sum_{i=1}^\ell \mu_i\lambda_i =0$ without any restriction for $\mu_{f_\mu}$.

Finally, given a specific values for $\lambda_1, \lambda_2, \cdots, \lambda_{\ell}$, with $\lambda_{f_\lambda} =1$, our problem consist of counting how many solutions $(\mu_1,\mu_2,\cdots,\mu_{m_a} )$ the equation $1+\sum_{i=1}^\ell \mu_i\lambda_ie =0$ has, without any restriction for $\mu_{f_\mu}$. It is easy to see that this value is $q^{\ell-1}$.

Doing the above account for all the $\dd \frac{q^{\ell}-1}{q-1}$ possibilities when you choose the specific values for $\lambda_1, \lambda_2, \cdots, \lambda_{\ell}$ we reach the statement.
\end{proof}

\section{Kronecker product construction of uniformly packed codes}

The following theorem describes the explicit construction of infinite
family of $q$-ary linear uniformly packed codes (in the wide sense) with
fixed covering radius $\rho$, where $q$ is any prime power, and
where $\rho \geq 2$ is an arbitrary natural number. The interesting fact
here is that these codes are not completely regular.

Recall that a trivial $q$-ary repetition $[n, 1, n]_q$-code is a
perfect code, if and only if $q=2$ and $n$ is odd.

\begin{theo}\label{theo:4.3}
Let $C_A$ and $C_B$ be two linear codes: the  repetition $[n_a,1,n_a]_q$-code $C_A$ of length
$n_a \geq 3$ and the $q$-ary perfect Hamming
$[n_b,k_b,3]_q$-code $C_B$ of length $n_b=(q^m-1)/(q-1) \geq q+1$, where $n_a \leq n_b$.
Let $A$
(respectively, $B$) be a parity check matrix of code $C_A$
(respectively, $C_B$). Then the matrix $H = A \otimes B$, the
Kronecker product of $A$ and $B$, is a parity check matrix of a
$q$-ary uniformly packed (in the wide sense) $[n, k, d]_q$-code $C$
with covering radius $\rho$, where
\EQ\label{eq:4.3.1}
n = n_a {\cdot} n_b,~~k = n - m{\cdot}(n_a-1),~~d = 3,~~\rho = n_a-1.
\EN
Furthermore, code $C$ is not completely regular with an exception for
the case $q=2$ and $n_b=3$.
\end{theo}

\begin{proof}
Let $\ba_i$ (respectively, $\bb_j$) denotes the $i$-th column of
$A$ (respectively, the $j$-th column of $B$).

Remark that
$A$ is a $(n_a-1 \times n_a)$-matrix:
\[
A ~=~ \left[
\begin{array}{ccccc|c}
&1~&0~&0~~\cdots~ &0~&-\,1~\\
&0~&1~&0~~\cdots~ &0~&-\,1~\\
&0~&0~&1~~\cdots~ &0~&-\,1~\\
&\cdot~&\cdot~&\cdot~~\cdots~ &\cdot~&~\cdot~\cdot~\\
&0~&0~&0~~\cdots~ &1~&-\,1~\\
\end{array}\right].
\]
Hence, the matrix $H = A \otimes B$ has a very simple structure:
\[
H ~=~ \left[
\begin{array}{ccccc|c}
&B~&0~&0~~\cdots~ &0~&-\,B~\\
&0~&B~&0~~\cdots~ &0~&-\,B~\\
&0~&0~&B~~\cdots~ &0~&-\,B~\\
&\cdot~&\cdot~&\cdot~~\cdots~ &\cdot~&~\cdot~\cdot~\\
&0~&0~&0~~\cdots~ &B~&-\,B~\\
\end{array}\right],
\]
where $0$ denotes the zero matrix of size $m\times n_b$.

Any $q$-ary vector $\bx$ of length $n-k=m{\cdot}(n_a-1)$
can be presented as follows:~ $\bx = (\bx_1\,|\,\ldots \,|\,\bx_{n_a-1})$
where $\bx_i$ is a $q$-ary vector of length $m$ for any $i=1,\ldots,n_a-1$.
Matrix $B$ contains as columns, up to multiplicative scalar, any vector
over $\F_q$ of length $n_b-k_b = m$. Hence for any
$\bx_i$,~$i=1,\ldots,n_a-1$ there is a column $\bb_{j_i}$ of $B$ such that
$\bx_i^t = \xi_i \bb_{j_i}$ for some $\xi_i \in \F_q^{*}$. Since $\ba_i \otimes \bb_j$
is a column of $H$, and since $\bx^t$ can be written as $\bx^t= \sum_{i=1}^{n_a-1} \xi_i\bb_{j_i}$,
we deduce that $\rho \leq n_a-1$. To see that $\rho \geq n_a-1$ it
is enough to choose as a vector $\bx$ a vector with all nonzero mutually
different component vectors $\bx_i$, ~$i=1,\ldots, n_a-1$. Such
a choice is possible, since $q^m - 1 \geq n_b \geq n_a$. We conclude that $\rho = n_a-1$.

Now we turn to the outer distance $s=s(C)$ of $C$ (i.e. the
number of different nonzero weights of codewords in
$C^{\perp}$).

Matrix $B$ is the parity check matrix of a Hamming code so, after
Lemma~\ref{lem:4.1a}, we conclude that all the nonzero linear
combinations of the rows in $A$ have the same weight $q^{m-1}$.

Now consider any linear combination over $\F_q$ of rows of $H$.
It is easy to see, by the shape of $H$ that the number of different
nonzero weights go from $2{\cdot}q^{m-1}$ until $n_a{\cdot}q^{m-1}$
so, the number of different nonzero values for the weight of the
codewords in the code $C^{\perp}$ generated by the matrix $H$
is equal to $n_a-1$. Hence, the outer distance $s(C)$
of $C$ is equal to $n_a-1$ and so, $\rho(C) = s(C)$. Now, using Lemma \ref{lem:2.3},
we conclude that the code $C$ is uniformly packed in the wide
sense, i.e. in the sense of \cite{Bas1}.

To finish the proof we have only to show that $C$ is not completely
regular, with only one exception: when $A$ is the trivial binary
repetition $[3,1,3]_2$-code which, at the same time, is the
trivial Hamming code of length $3$. But this last case (i.e. the case
$q=2$ and $n_b=3$) is included in Theorem \ref{theo:4.1}.
Hence we have only to show that in all other cases the code
$C$ is not completely regular.

Consider the next possible binary repetition code. When $n_a=4$ and
$\rho=3$ we have the repetition $[4,1,4]_2$-code $C_A$. Choose as the
code $C_B$ the binary Hamming $[7,4,3]_2$-code. We claim that the
resulting $[28, 19, 3]_2$-code $C$ (after applying Theorem~\ref{theo:4.3})
is not a completely regular code. Let $H= A \otimes B$, i.e. $H$ looks
as
\[
H ~=~ \left[
\begin{array}{ccccc}
&B~&0~&0~&B~\\
&0~&B~&0~&B~\\
&0~&0~&B~&B~\\
\end{array}\right],
\]
where $B$ is the following matrix:
\[
B ~=~ \left[
\begin{array}{cccccccc}
&1~&0~&0~&1~&0~&1~&~1~\\
&0~&1~&0~&1~&1~&0~&~1~\\
&0~&0~&1~&0~&1~&1~&~1~\\
\end{array}\right].
\]
Consider two different vectors $\bx_1$ and $\bx_2$ of weight
$2$, which belong to $C(2)$. Let
\[
\bx_1~=~(1000000|1000000|0000000|0000000)
\]
and
\[
\bx_2~=~(1000000|0100000|0000000|0000000).
\]
It is easy to see that both vectors $\bx_1$ and $\bx_2$ are from $C(2)$
and we obtain immediately the intersection numbers $c_2=4$ for $\bx_1$ and
$c_2=2$ for $\bx_2$. Thus, code $C$ is not completely regular. Clearly the
same contra-example works for $q=2$ and for larger values $n_b \geq 5$.

For the cases $q \geq 3$ these above contra-examples should be slightly modified.
For the smallest case $q=3$ and $n_b=3$ choose
as the code $C_B$ the Hamming $[4,2,3]_3$-code with parity check matrix
$B$ and let $A$ be a parity check matrix of the repetition $[3,1,3]_3$-code
$C_A$, where $\F_3=\{0,1,2\}$,
\[
B=\left[
\begin{array}{cccc}
1~&1~&1~&0\\
0~&1~&2~&1
\end{array}
\right]
~~\mbox{and}~~
A=\left[
\begin{array}{cccc}
&1~&0~&2\\
&0~&1~&2
\end{array}
\right]
\]
Take the following vectors $\bx_1$ and $\bx_2$
from $C(2)$:
\[
\bx_1~=~(1000|2000|0000)~~
\mbox{and}~~\bx_2~=~(1000|0100|0000).
\]
We obtain the intersection numbers $c_2=4$ for $\bx_1$ and $c_2=2$ for $\bx_2$. Hence, the resulting $[12,8,3]_3$-code $C$ is not completely regular. The same contra-example works for the rest of cases $q\geq 3$ and $n_b \geq 3$.

Now, the proof of the theorem is complete.
\end{proof}

\end{document}